%% file: main.tex
\documentclass{llncs}

\usepackage{graphicx}

\usepackage{hyperref}

\begin{document}
\title{Blockchain and Principles of Business Process Re-Engineering for Process Innovation}

\input{Abstract}
\input{Introduction}
\input{ConceptualF.tex}

\input{InsuranceCase}

\input{BPR}
\input{Discussion}

\input{Conclusion}
\bibliographystyle{plain}
\bibliography{Bibliography}
\end{document}

%% file: Abstract.tex
\titlerunning{Blockchain and Principles of Business Process Re-Engineering for Process Innovation}  
%
\author{Fredrik Milani \and Luciano Garc\'{\i}a-Ba\~nuelos}
\authorrunning{Fredrik Milani and Luciano Garc\'{\'i}a-Ba\~nuelos} 
%
\tocauthor{Fredrik Milani, Luciano Garc\'{\i}a-Ba\~nuelos}
\institute{University of Tartu, Liivi 2, 50409 Tartu, Estonia \\
E-mail \{milani, luciano.garcia\}@ut.ee}

\maketitle              

\begin{abstract}
Blockchain has emerged as one of the most promising and revolutionary technologies in the past years. Companies are exploring implementation of use cases in hope of significant gains in efficiencies. However, to achieve the impact hoped for, it is not sufficient to merely replace existing technologies. The current business processes must also be redesigned and innovated to enable realization of hoped for benefits. This conceptual paper provides a theoretical contribution on how blockchain technology and smart contracts potentially can, within the framework of the seven principles of business process re-engineering (BPR), enable process innovations. In this paper, we analyze the BPR principles in light of their applicability to blockchain-based solutions. We find these principles to be applicable and helpful in understanding how blockchain technology could enable transformational redesign of current processes. However, the viewpoint taken, should be expanded from intra- to inter-organizational processes operating within an ecosystem of separate organizational entities. In such a blockchain powered ecosystem, smart contracts take on a pivotal role, both as repositories of data and executioner of activities.

\keywords{Blockchain, Business Process Management, Process Innovation, BPR, Research Challenges}
\end{abstract}

%% file: Introduction.tex
\section{Introduction}
\label{Introduction}

Blockchain is speculated to radically transform how companies conduct their business and thereby, their processes. Investment and attention to this technology has increased significantly the past years. It has been called the "next Internet" and described to impact business processes with a similar magnitude as process automation did for manufacturing and service industries in the 1990-ties. Although is often frowned upon due to its connection with the volatile and speculative nature of Bitcoin, many companies are interested in the underlying technology and investing in better understanding its usage beyond cryptocurrencies.

At the core of blockchain, lies the concept of a distributed ledger. In essence, distributed ledgers stores its records, such as transactions, on a public ledger. This ledger is not hosted at one centralized and trusted institution but distributed. Every participating node therefore, holds a full copy of the ledger. Any change needs to be agreed by the majority of the nodes before it is accepted and populated across all ledgers. As such, verification is not dependent on one central party but on the participants. Once a transaction is recorded in the chain, it is immutable. The benefits of blockchain to enable proof of identity, enable cross-border collaboration, secure and immutable records of asset transfer or movement of goods, has been explored by both incumbents and start-ups.

However, we have learned from history that introduction of new business enabling technologies does not automatically result in ability to capitalize on innovations emerging therefrom. For instance, the productivity of factories only improved marginally with the introduction of electric motors. It would take about 30 years before the productivity improved significantly because factory managers simply replaced the steam engines with electric ones. It was not until the processes were changed that the potential benefits of the new technology could be fully utilized \cite{brynjolfsson2014second}. The same argument can be made for blockchain. If these technologies are simply used to replace existing solutions, the added value will be noticeable but comparatively marginal. However, if business processes are redesigned with consideration to the enabling powers of blockchain technology, significant gains in efficiency can be achieved. Therefore, the business process management perspective is crucial for successful integration of blockchain based solutions in companies \cite{milani2016blockchain}.

Although blockchain technology has not yet matured, it has gained traction and hold potential to impact BPM research \cite{mendling2018blockchains}. In light of this context, it is worth examining how experiences with transformational process improvements \cite{dumas2013fundamentals} can potentially be applied to blockchain. In this paper, we focus on its potential impact to enable creation of novel business processes. To this end, the main research question of this paper is as follows. "\textit{How can blockchain technology impact business processes redesign?}" In discussing this questions, we focus on exploring the principles of business process re-engineering (BPR) \cite{hammer1990reengineering} and how these could potentially guide and be used to better understand improvement opportunities when innovating processes on blockchain platforms. We use an example inspired by a prototype showcased by IBM \footnote{https://www.ibm.com/blogs/insights-on-business/government/blockchain-asset-registration/}. This case concerns the manufacturing and insuring of a customized high-end guitar. Then we consider how this process could be innovated, guided by the seven principles of BPR.

The remainder of the paper is structured as follows. Section \ref{sec:ConceptualF} introduces the conceptual foundations of blockchain and smart contracts. Section \ref{sec:InsuranceCase} introduces the example case of the stolen high-end guitar. In Section \ref{sec:BPR}, we  examine each and one of the seven principles of BPR followed by a discussion in Section \ref{sec:Discussion}. Finally, we conclude the paper and outline possible venues for future work in Section \ref{sec:Conclusion}.

%% file: ConceptualF.tex
\section{Conceptual Foundations}
\label{sec:ConceptualF}

Blockchain can be understood as a write-only, shared database \cite{yli2016current}. In contrast to conventional databases, a blockchain is intended to store information in an append-only fashion, disallowing any change to already stored information. To this end, the blockchain is implemented using a sophisticated consensus algorithm that run over a peer-to-peer network. The consensus algorithm is used to decide, as a result of executing transactions, what information to add into the blockchain. The information added by transactions is gathered into blocks connected to together as a chain using crytographical pointers (hence the term blockchain). These pointers makes it virtually impossible, or else prohibitively computationally expensive, to change the contents already stored in the blockchain.

Blockchain was initially introduced to enable a digital currency called Bitocoin \cite{nakamoto2008bitcoin}. In the case of Bitcoin, the intent of transactions, e.g. transfers of cryptocurrency between digital wallets, is encoded using a fairly simple scripting language. The possibility of having a digital ledger, i.e. blockchain with programming capabilities, intersects with Nick Szabo’s earlier idea of a smart contract \cite{szabo1997formalizing}. However, the simplicity of Bitcoin’s scripting language has been found to have limitations. In addressing some of these limitations, the so-called second-generation of blockchains have evolved. These incorporate more expressive programming capabilities. Although the use of smart contracts is still under development, from both commercial implementations and legal perspective, the provision of more expressive programming capabilities has opened other possibilities, which we describe further below.

One of the main use cases for smart contracts is as a digital asset. Blockchains and smart contracts have been found to be a convenient way to represent the legal rights over assets. This is enabled by the concept referred to as tokenization. Both tangible assets such as diamonds \cite{underwood2016blockchain} or real estate \cite{spielman2016blockchain} as well as intangible ones such as cryptocurrencies \cite{peters2015trends} can be represented and captured with tokenization. In this context, a smart contract is intended for storing information about the asset (e.g. cryptographic hash values to represent digital information such as photos, serial numbers and additional information) and information to bind the asset with its owner.

Smart contracts can also be used for encoding business logic. The expressive power of blockchain programming languages have been found suitable for encoding the logic behind business processes \cite{luu2016making}. A smart contract can be associated with a running instance of a business process, storing case related information and encoding the control flow and business rules thereof.

Furthermore, smart contracts can be used for identity validation. Smart contracts are used in keeping track of the participation of stakeholders on business transactions and in allowing access to only authorized stakeholders at any given step of the process. The validation of validation is performed by means of digital signatures that are controlled by cryptographic protocols and verified within smart contracts.

%% file: InsuranceCase.tex
\section{Process Re-engineering of an Insurance Case}
\label{sec:InsuranceCase}

In this Section, we first present the illustrating example we use for discussing each one of the principles of BPR. We also broadly outline the development of context within which BPR has been proposed. Following this, we introduce the viewpoints made possible with the enabling technology of blockchain and smart contracts.

\subsection{Introduction of the Insurance Case}

\begin{figure}[htb!]
	\centering
    \includegraphics[width=\textwidth]{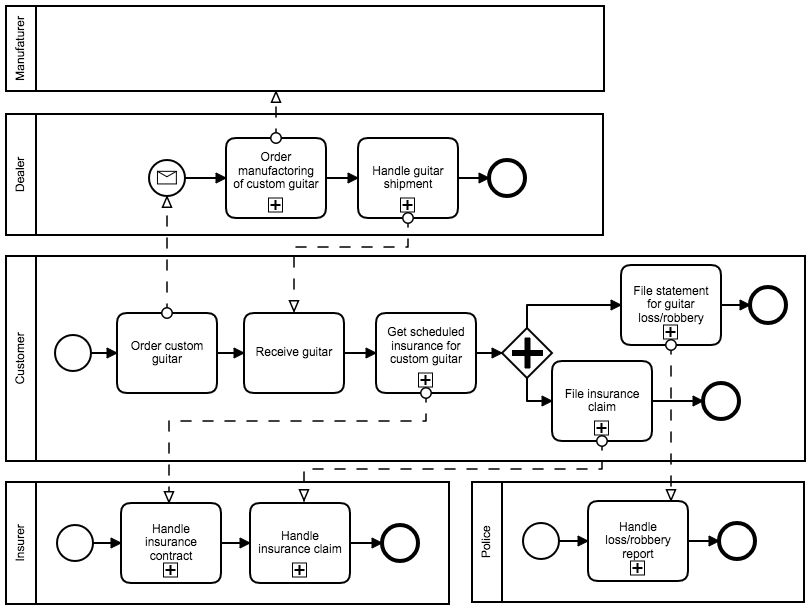}
	\caption{Main Process of the Insurance Case}
	\label{fig:mainprocess}
\end{figure}

The insurance case used as example is based on an IBM case \footnote{https://www.ibm.com/blogs/insights-on-business/government/blockchain-asset-registration/}. In this example, the customer orders a custom-made guitar from a dealer. The dealer manages the order and forwards it to the manufacturer who produces and ships the guitar to the dealer. Following this, the dealer makes sure the guitar gets to the customer. As it is an expensive guitar, the customer takes out an insurance. We also assume that the guitar gets stolen at some point. The customer files a police reports which is handled by the police department. Then the owner files a claim with the insurance company. The claim is handled and eventually approved. Figure \ref{fig:mainprocess} depicts the main process of this insurance case. 
We can also assume that each the guitar, being high-end custom-made one, is tagged with a unique id and possibly has a small gps tracker embedded in the neck of the guitar. We use this simple example to illustrate the applicability of the seven principles of business process re-engineering \cite{hammer1990reengineering}. In so doing, we will introduce slightly more complex situations and variations.

\subsection{Evolution of the Context of BRP}
Michael Hammer began his famous paper by stating that "managers can release the real power of computers by challenging centuries-old notions about work" \cite{hammer1990reengineering}. At the core, the idea is that processes can be changed to achieve significant improvements but only of they are supported by information technology. A mere replacement of existing technology with a new one does not yield drastic improvements. This notion is supported by others such as \cite{brynjolfsson2014second}. Hammer's scope was to use information technology within the boundaries of an organization to change the processes. The example of Ford's accounts payable process illustrate this point. The different divisions, once operating as separate entities, were united with the support of an common mainframe system.

Extending this idea, Champy \cite{champy2005x} took the concept of re-thinking how work is done across organizational boundaries. He defined four levels of re-engineering. At the first level, a company re-designs its own processes. This first level corresponds with Hammer's ideas. At the second level, the company redesign processes in collaboration with one type (supplier, customer, or partner) of external business organization. The third level is essentially the same but with the exception of the re-design taking place with two types of external businesses (for instance supplier and customer). At the highest level (the fourth level), the processes of the company are redesigned in close collaboration with all the types of external business organizations (supplier, customer, and partner). The scope of the fourth level includes the most important adjacent external companies. In short, Hammer began with the company in focus and Champy extended the scope to include the most important "neighbors".

\subsection{The Context within the Framework of Blockchain and Smart Contracts}

The scope of both Hammer and Champy was aligned with what was feasible to achieve at that time considering the capabilities of information technology. However, blockchain has the potential to break the technological "glass ceiling". This has two implications worth considering when challenging decade-old notions about work.

The first regards an ecosystem of entities. It is now possible to consider processes across an ecosystem of organizations. This is in contrast to processes being confined to a company or its immediate neighbors. In the example presented above, the ecosystem includes manufacturer, dealership, insurance company, police, and in particular, the customer. Commonly, collaboration across an ecosystem requires a trusted third party or an integration platform. Both of these alternatives incur costs that outweigh the benefits, in particular if transaction volumes are below a certain threshold. However, with the blockchain technology it is possible to extent process redesign from an intra-company and inter-company to an ecosystem. Such an ecosystem includes for instance, governmental agencies, private corporations at different points of the supply chain, non-governmental agencies, and perhaps regulatory institutions.

The second is a shift in focus is from company to customer. Process redesign has traditionally considered the internal processes of a company. Although this has been extended across organizational boundaries, it is still confined to an overall process such as a supply chain flowing through several companies. Even when processes are improved for the benefit of the customer, it is from the perspective of the company. With blockchain technologies, it will be possible to take the viewpoint of the customer. For instance, in the insurance case presented above, each organization such as the dealer, police, and insurance company has optimized their processes and made it easy for the customer to interact with them. The customer might have an excellent and seamless experience with each organization but still, they have to take the reports from one entity to another. While the touch points are optimized, the customer process (journey) is not.

The principles of business process re-engineering aimed at using the power of information technology to rethink work. Again we are presented with information technology that can push the frontier of process innovation further \cite{milani2016blockchain}. In this context, we examine the applicability of the seven principles of BPR in the setting of the insurance case when engined with blockchain technology and smart contracts.

%% file: BPR.tex
\section{Applying the Principles of BRP}
\label{sec:BPR}

BPR projects have been critically reviewed. Studies show that BPR projects have failed \cite{kleiner2000revisiting} or delivered results less than expected \cite{grover2000process} while others conclude positive results \cite{ozcelik2010business}. The failures does not seem to be attributed to the BPR principles but rather to factors such as lack of partnership and communication between for instance IT and other departments of the company \cite{hammer1993re}.
At the heart of BPR lies the idea of breaking foundational patterns of thinking underlying existing business processes. Only a transformational restructuring where old assumptions are challenged, can lead to significant gains in performance. The enabling factor that allow for processes to be restructured in a revolutionary way is technology. In this light, blockchain has been heralded as the technology that can uproot conventional process structures. Therefore, we examine each and one of the principles outlined by Hammer \cite{hammer1990reengineering} to understand how current processes could be innovated if supported by blockchain technology with smart contracts.

\subsection{Principle 1 - Organize around outcomes, not tasks}

The first principle concerns the division of work for one process instance. Hammer uses the example of an electronics company where different departments performed the five different required steps. One department was responsible for determining customer requirements, another translated them into codes followed by a third who disseminated the information to different plants and warehouses. Then another division assembled the product and finally, a fifth one delivered and installed the equipment. This way of organizing around tasks in a sequential manner, led to errors, rework, and delays. Oftentimes, excessive efforts were put in overseeing and tracking progress of a case.

The insurance example follows a similar pattern. Let us assume that the guitar gets stolen and the insurance company covers the costs of replacing it. Here, the owner has to take the case to the police (first division), and then wait for the report. Once the report is generated and received, the owner will take it to the insurance company (second division). Finally, the insurance company will send the funds to the dealership (third division) once a new guitar has been ordered. The customer is made responsible for overseeing and tacking the progress, many times at the cost of time and frustration. Should the police recover the guitar and apprehend the thief, the insurance company would file a claim of compensation as well. That would involve the police, the court (fourth division), and the insurance company. If the thief is found guilty but unable to pay, the enforcement authority (fifth division) will be involved to confiscate assets of the thief to be sold. In short, taking the ecosystem and the customer perspective, be it the buyer of the guitar or the insurance company trying to get compensated, we see similarities.

In redesigning the processes of the electronics company, the work was organized around outcome (not tasks) by means of "customer service representative". In a similar manner, the insurance company used in Hammer's paper, "case managers" responsible for the application-to-policy process, were introduced. Such new roles were made possible due to shared databases as the case managers could access and record all the data required. Blockchain technology enables replacing the case managers with the concept of smart contracts. A set of connected smart contracts can become the new "case managers". Note that it will require several different types of smart contracts, each created when triggered by an event, and linked to each other to collectively function as the case manager.

In the example of this paper, a smart contract is created when a new guitar is ordered (smart contract for provenance). As it progresses through manufacturing, it gets populated with additional data. Once the guitar is completed and full payment made, the customer is registered as the owner (smart contract for asset registry). When the guitar is stolen, a new smart contract is used to file a claim with the police. The police issues a report which is taken as input by the insurance company. Once the guitar has been reported stolen, it is marked as stolen in the smart contract (asset registry), rending the guitar very difficult to fence. The claim (using smart contract for registry and report) is investigated and approved. As such, a smart contract for payout is created which the owner can use to order and pay for a new replacement guitar. Tracing and tracking progress is inherently present with the smart contracts. Such a solution also allow for provenance to ensure that the guitar is manufactured according to the quality required.  Although this example is trivial, it illustrates the potentiality of smart contracts to replace case managers. The role once occupied by a humans, can now be replaced with a set of smart contracts. These contracts will make no mistakes, will not fluctuate in their performance, will not be complicit in fraud, and can manage very large volumes of transactions.

\subsection{Principle 2 - Have those who use the output of the process perform the process}

Hammer noted that organizations created specialized departments to manage specific processes. He uses the example of internal departments being as customers when turning to the central purchasing departments for needed items. For expensive and unique items, such a solution is good but for most ordinary items such as pencils, the cost of managing the process is higher than the cost of the pencils. The second principles states that the internal departments who use the output of the purchasing process (pencils) should perform the process (place the order) themselves.

In a way, this principle pushes the work to the consumer of the output. Today, this principle has matured and is visible in the concept of self-service. For instance, rather than a clerk at an insurance company registering the data, the customer does the work by filling in the data via an on-line portal. In supermarkets, customers can themselves scan their products and pay at self-service stations. With such solutions, both customers and companies are benefiting. The companies reduce costs and the customers find it more convenient (filling in the data at a time of their own choosing or not needing to wait in line to buy groceries).

In an ecosystem as in the example used in this paper, we see a number of specialized departments (entities) who perform the process. The output is then used by another entity. For instance, the police report is used by the insurance company and the payout is used by the dealer. Across different entities, it is not possible to apply this principle. For instance, it is not reasonable to have the guitar owner (user of the claim assessment process), do the work. Nor is it legally sound for the insurance company to generate a police report. However, it is in the guitar owner's interest to have all these sub-processes performed so to receive compensation for the stolen guitar. In this regard and in light of self-service, the guitar owner can initiate the overall process of getting compensation. This in contrast to  initiate a set of sub-processes. For instance, the guitar owner has to go to the police, file and receive a police report, take it to the insurance company, file a claim, and regularly check the progress. Instead, with the use of smart contracts, the guitar owner can trigger the overall process by initiating a smart contract for stolen guitar. In this process, the owner links the smart contract proving ownership to the insurance smart contract. Following this, the smart contracts can take the process further. The smart contracts become the extended arm of the customer. They represent the customer (who uses the output) and drives (performs) the process on his or her behalf. As such, the smart contracts initiate the sub-process with for instance the police department.

\subsection{Principle 3 - Subsume information-processing work into the real work that produces the information}

Yet another inherited rule born of the idea of specialized division of labor, is the separation of data collection and the processing of that same data. Hammer noted that companies had structures were one division collects and records the data while another actually uses the data. In the Ford example, the accounts payable department collected the information from two other divisions and reconciled the data with that provided from the vendors. In the redesigned process, the goods receiving department used the shared database to process the information rather than just entering and forwarding it. As such, the processing of the information was subsumed with producing the data.

The recording and production of data can be subsumed in an automated manner with the concept of smart contracts. A smart contract can be designed to have two main components tightly linked to each other. The first is a registry of data. For instance, the owner buys a guitar tagged with an unique id. An ownership (asset) smart contract is created where the id of the guitar and the owner's personal data is recorded. As such, the smart contract functions as a depository of data (ownership). On the other hand, smart contracts also include automated execution of business rules. This aspect allows smart contracts to execute a process or in other words, processing information. For instance, a smart contract for a certain asset, holds information about the guitar (by means of a unique id) and its owner. Should the current owner sell the guitar, the smart contract needs to be updated. As such, the seller and buyer updates the info. The smart contract will execute a set of related automated business rules (information-processing work for verification and transfer of ownership) and record the update in its depository of data (work that produce data).

Furthermore, other related processes could be triggered as well. For instance, when the guitar changes owner, the previous owner may wish to cancel his or her insurance. In an ecosystem enabled by blockchain, one transaction can, such as when change of ownership where data is updated, automatically trigger cancellation of existing insurance. As such, the smart contacts not only capture new data (change of ownership), processes that information (update the owner in its depository of data), but also set in motion related processes where information is updated and processed. Again, these triggered processes would in most cases subsume the information-processing into the real work.

\subsection{Principle 4 - Treat geographically dispersed resources as though they were centralized}
Hammer noted that with the advances in information technology, dispersed geographical locations could be treated as if they were all located at the same place. He uses the example of Hewlett-Packard whose manufacturing units had separate purchasing departments. By using a shared database, all could access the same data and become "as one". This enabled them to use the volume of orders to negotiate better agreements with vendors. 

In our example, the organizational unit are geographically dispersed. However, with blockchain, they can be treated as if they were centralized. From an ecosystem and customer viewpoint, a blockchain based solution enables considering the whole process as taking place under one roof. The blockchain would integrate all the different entities. The family of smart contracts would correspond to the shared database referred to by Hammer. Furthermore, the customer would perhaps only have one portal or interface for the whole process. Rather than visiting the police department, then the insurance company, or make re-visits for complementary information or receiving update on the progress, the customer uses the blockchain based solution to conduct all activities, receive notifications on updates, progress report, and information (notification) on when and what further action is required. In other words, the customer gets the same access as the "case manager" in Hammer's example.

\subsection{Principle 5 - Link parallel activities instead of integrating their results}

Hammer illustrates that parallel processing takes shape in two different ways. One is when separate divisions perform the same activity. The other is when different divisions produce outcomes that in the end must be integrated. The idea is to link these processes rather than integrating their end results. Hammer clarifies this principle by suggesting to "coordinate parallel function during the process - not after it's completed" \cite{hammer1990reengineering}. To remedy this, Hammer propose using communication networks, shared databases, and other means by which coordination between processes can happen during the process execution. At the end of the day, the need of inter-process communication by means of technology is proposed.

In the case of the stolen guitar, the owner has to secure a police report and take the outcome of this process to the insurance company. With blockchain technology and smart contracts, it is possible to link parallel processes across separate entities within an ecosystem. For instance, the owner of the stolen guitar can initiate an insurance claim that triggers both the generation of the police report as well as the processes within the insurance company. Assuming that most of the processing is automated, the claim can be processed while the police report is being generated. In a similar manner, with the claim approved, a smart contract for claiming financial compensation can be created. If the thief is caught, the claim is already in place. This allows insurance companies to secure that their interest is always represented in court. In regards to inter-process communication, smart contracts could assume the role of facilitating inter-process communication. Considering that in an ecosystem, different entities exist, the communication is often carried out via "emails" (corresponding to movement of papers in for instance the Ford case). With smart contracts at the core of connected ecosystem, communication between different entities is automated, efficient, and error-free.

\subsection{Principle 6 - Put the decision point where the work is performed, and build control into the process}

The sixth principle of BPR concerns cutting unnecessary controls and checks in a process. In the examples used by Hammer, work is conducted by certain resources whereas the decisions are taken by others. For instance, accountants, auditors, managers, supervisors regularly check and monitor the work of others. By empowering those closest to the activity and who do the work with decision authority, processes can be improved.

In an ecosystem, traces of this old pattern remains but in different form. In the case of the stolen guitar, the insurance company examines data about the owner, confirms the coverage and terms, review of the police report, and if this might be a fraudulent claim. However, these decisions are not taken where the work is done. The data about the owner is entered at a different point, the police report at another point, and the claim at yet another point. With the concept of smart contracts, data is verified when entered (where the work is conducted). The consensus algorithm that accepts a smart contract does the job of taking decisions about validity. When the customer buys a guitar, he or she is entered as the verified owner (smart contract for asset registry). If the guitar would be sold, the new owner is registered on the smart contract. Similarly, the controls can be built in the smart contracts. For instance, most straightforward cases can be automatically processed but given a set of business rules, certain cases are filtered out and sent for closer review.

\subsection{Principle 7 - Capture information once and at the source}

Hammer points out that when information is difficult to transmit, it makes sense to capture it several times and at different locations. However, when it can be distributed easily with for instance shared database structures, it no longer makes sense. Hence, the principle of capturing information once and at the source. In our example, the owner has to show proof of purchase, police report, and fill in information for the insurance claim. In many countries, these activities are still paper based. In each case, as the information is coming from outside of the organization, it has to be entered into the system. The same data is entered in the information systems of the police department as well as in the systems of the instance insurance company. Such data entry activities are costly, time consuming, error prone, and carries with it the risk of fraud requiring examination of authenticity of documents.

In an ecosystem, the concept of smart contracts can enable capturing information once and at the source. The data is captured once in the smart contract and by the entity initiating it (source). Updated data is also only captured once and at the source. Furthermore, blockchain technology ensures authenticity, traceability of data, and integration with companies own information systems.

%% file: Discussion.tex
\section{Discussion}
\label{sec:Discussion}

\begin{figure}[htb!]
	\centering
    \includegraphics[width=\textwidth]{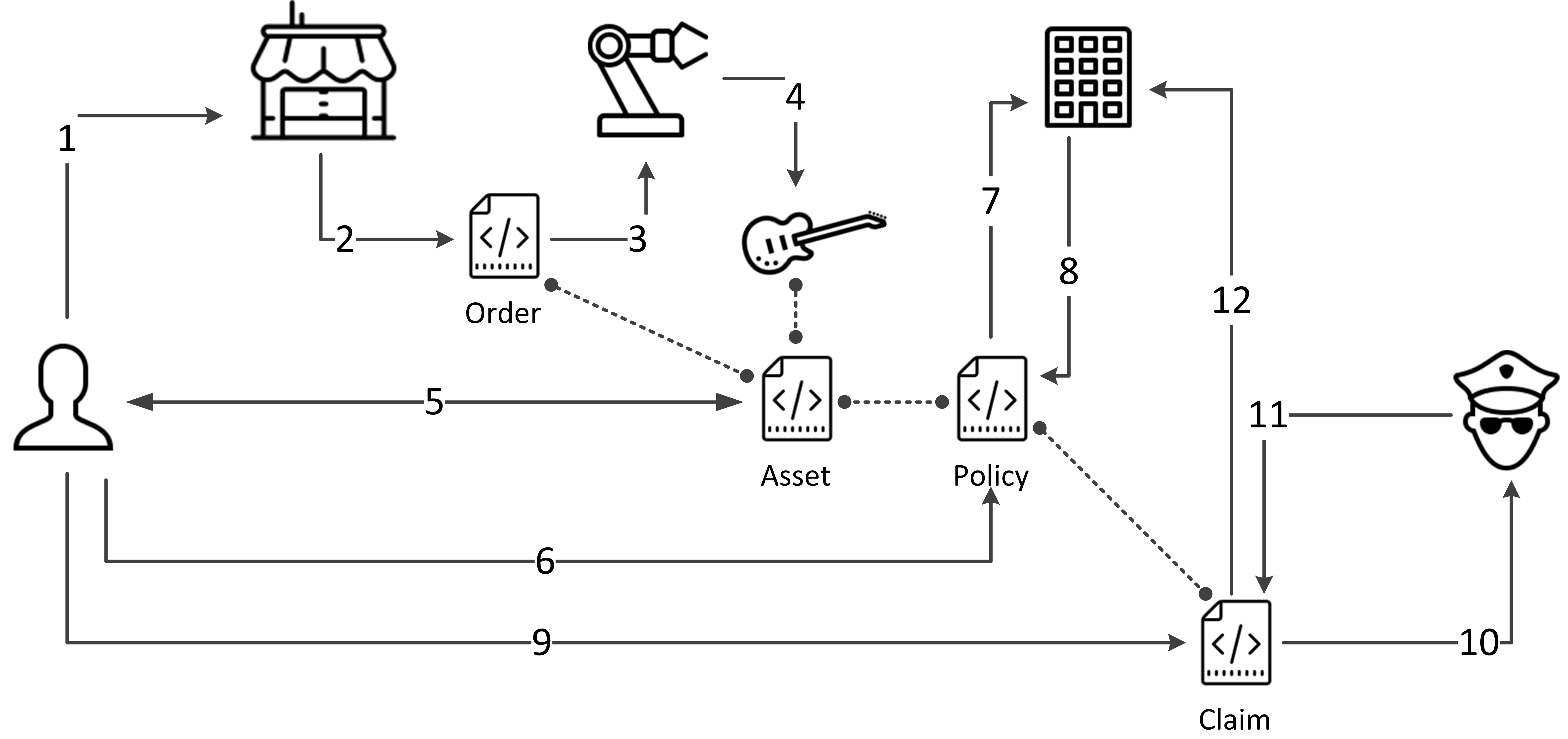}
	\caption{Blockchain based solution of the Insurance Case}
	\label{fig:Tobe.png}
\end{figure}

Blockchain technology provides the fundamental platform needed to connect separate entities within an ecosystem. We have examined how blockchain could potentially contribute to the redesign of business processes in light of BPR principles. In this exercise, smart contracts carry a pivotal role. Such contracts are the necessary component that enable execution of inter-organizational processes. In the example of this paper, we have modeled the as-is process in a very simple way. However, modeling an overview of the to-be state, was challenging. Figure \ref{fig:Tobe.png} gives a very simple overview of a potential solution but falls short of being a business process.
In a potential solution upon blockchain technology, the customer makes an order via dealership. This is captured with a smart contract that is used when manufacturing the guitar. Once completed, a smart contract is created to denote who the owner is. The contract for ownership is also connected to the "order" contract. The customer can then use the contract to insure the guitar. Should the guitar get stolen, the owner triggers a smart contract for stolen property. This contract, connected to the asset contract (ownership), is used as input for the claim assessment and enriched with for instance the police report. To keep the overview simple, we did not capture the insurance payout or the processes related to the claim of the insurance company against the thief.

In trying to model the potential to-be process, challenges arise. One is the complexity of the models. If the above process would be fully represented in a model, it would be a very complex model. Some form of process architecture would be helpful in modeling ecosystems. Another concern is the perspective. From what or whose perspective(s) should one model such processes? The perspective of the customer seems to be central. While customers are often modeled as a black or gray box \cite{dumas2013fundamentals}, they need to be brought into light much more. However, the core of the process lies smart contracts as "repositories of data" and "automated business rule execution". As such, the data perspective gains more importance as does the business rules. It does not seem that the current BPMN standard can adequately capture inter-organizational processes powered by blockchain technology. As such, there is a need to find a better way to model novel inter-organizational processes in an ecosystem \cite{mendling2018blockchains}.

Hammer proposed re-defining the job description and the activities performed by the roles. For instance, the first principle calls for creation of case managers or customer representatives. The second principle calls for empowering those who use the output to perform the process while the third principle seeks to subsume processing of data with production of data. The sixth principle suggest taking decisions where the activity is performed. These are all, in one way or another, redefining the role of process participants. In an ecosystem on the blockchain, this traditionally "human" role can be automated with smart contracts. Smart contracts can become the "case manager" of the customer (guitar owner). At the same time, these contracts can be the aid of the customer to drive the process. In this execution of the process, creation of data is subsumed with its processing and decisions are taken where the activity is performed. Furthermore, smart contracts can become the a conflict resolution point. Given the business rules, disputes are settled in an impartial and speedy manner. Smart contracts hold the data, keep a track of progress, and trigger activities when needed. In short, what is done manually by human resources in an inter-organizational process today, is done by smart contracts in an ecosystem on a blockchain. To enable this, smart contracts need to support two functions. The first is to act as a depository of data. Contracts must be able to create, read, update but perhaps not delete data. The other function of automated execution of business rules. These functions enable smart contracts to replace the humans in inter-organizational processes.

Most current use cases of blockchain aim at solving a specific problem. For instance, the problem of lack of monitoring within supply chain is used as an example in \cite{weber2016untrusted} or transparency \cite{francisco2018supply}. Within financial industries, banks explore blockchain for clearing and settlement of financial assets \cite{peters2016understanding}. Such a viewpoint might carry the risk of emulating the current as-is processes rather than redesigning them. For instance, a supply chain process might have issues with lack of transparency. This is due to lack of technology for monitoring the process across the different steps. By developing a blockchain based solution to digitally connect the various monitoring steps, even if it is enhanced by sensors along the chain, is replacing existing technology (manual work such as inspection and validation of documents) with an automated one. This is perhaps not surprising considering that current redesign heuristics is based on recurring patterns \cite{vanwersch2016critical} and we simply do not yet have such patterns to learn from. If this holds true, processes are not truly redesigned but perhaps essentially replacing electronic data interchange (EDI) with smart contracts. With a redesign perspective, it might be possible to see opportunities where functionality is "moved" from the IT systems of the companies to the smart contracts and the ecosystem. Another example might be to open up a "market" within the ecosystem. In the example of this paper, the ecosystem could be used by the customer to request quotes from insurance companies. Insurance companies would give their offer and the guitar owner would choose the most suitable insurance policy.

%% file: Conclusion.tex
\section{Conclusion}
\label{sec:Conclusion}

Although blockchain technology has numerous challenges  \cite{yli2016current} that must be solved, its potential power has attracted much attention. The promise it holds is particularly relevant for the BPM community. If blockchain technology replaces existing processes (as some preliminary signs might indicate), the potential gains will be limited. It is therefore, important to consider how blockchain and smart contracts can enable process redesign. In this regard, the BPM community is in a strong position to offer assistance via research.

This paper set out to explore "how blockchain technology can impact business process redesign". In so doing, we examined and analyzed the applicability of the seven principles of BPR as introduced by Hammer {\cite{hammer1990reengineering}}. We note that the blockchain technology enable separate organizational entities to interact as if existing within an ecosystem. We also found that smart contracts can potentially play a pivotal role in achieving efficiency gains. Their capability to act as repository of data and executor of activities, enable rethinking processes by for instance questioning where functionality should be placed (within a company IT system or in the ecosystem). Furthermore, such contracts enable automating manual activities such as data entry and dispute settlement.

Process redesign is not a trivial task. When analyzing processes, models play an important role. Limitations in modeling inter-organizational processes in a blockchain based ecosystem with links to smart contracts, poses a challenge. The role of smart contracts in enabling opening of market spaces in an ecosystem, and its capabilities of taking on functionalities traditionally within IT systems, are equally interesting questions. These considerations and limitations open interesting venues for future work.

%% file: main.bbl
\begin{thebibliography}{10}

\bibitem{brynjolfsson2014second}
Erik Brynjolfsson and Andrew McAfee.
\newblock {\em The second machine age: Work, progress, and prosperity in a time
  of brilliant technologies}.
\newblock WW Norton \& Company, 2014.

\bibitem{champy2005x}
James Champy.
\newblock {\em X-Engineering the corporation: Reinventing your business in the
  digital age}.
\newblock Recording for the Blind \& Dyslexic, 2005.

\bibitem{dumas2013fundamentals}
Marlon Dumas, Marcello La~Rosa, Jan Mendling, Hajo~A Reijers, et~al.
\newblock {\em Fundamentals of business process management}, volume~1.
\newblock Springer, 2013.

\bibitem{francisco2018supply}
Kristoffer Francisco and David Swanson.
\newblock The supply chain has no clothes: Technology adoption of blockchain
  for supply chain transparency.
\newblock {\em Logistics}, 2(1):2, 2018.

\bibitem{grover2000process}
Varun Grover and William~J Kettinger.
\newblock {\em Process think: winning perspectives for business change in the
  information age}.
\newblock IGI Global, 2000.

\bibitem{hammer1993re}
M~Hammer and J~Champy.
\newblock Re-engineering the corporation: A manifesto for business revolution.
\newblock {\em New York}, 1993.

\bibitem{hammer1990reengineering}
Michael Hammer.
\newblock Reengineering work: don't automate, obliterate.
\newblock {\em Harvard business review}, 68(4):104--112, 1990.

\bibitem{kleiner2000revisiting}
Art Kleiner.
\newblock Revisiting reengineering.
\newblock {\em Strategy+ Business}, 20(3):27--31, 2000.

\bibitem{luu2016making}
Loi Luu, Duc-Hiep Chu, Hrishi Olickel, Prateek Saxena, and Aquinas Hobor.
\newblock Making smart contracts smarter.
\newblock In {\em Proceedings of the 2016 ACM SIGSAC Conference on Computer and
  Communications Security}, pages 254--269. ACM, 2016.

\bibitem{mendling2018blockchains}
Jan Mendling, Ingo Weber, Wil Van~Der Aalst, Jan~Vom Brocke, Cristina
  Cabanillas, Florian Daniel, S{\o}ren Debois, Claudio~Di Ciccio, Marlon Dumas,
  Schahram Dustdar, et~al.
\newblock Blockchains for business process management-challenges and
  opportunities.
\newblock {\em ACM Transactions on Management Information Systems (TMIS)},
  9(1):4, 2018.

\bibitem{milani2016blockchain}
Fredrik Milani, Luciano Garc{\'\i}a-Ba{\~n}uelos, and Marlon Dumas.
\newblock Blockchain and business process improvement.
\newblock {\em BPTrends newsletter (October 2016)}, 2016.

\bibitem{nakamoto2008bitcoin}
Satoshi Nakamoto.
\newblock Bitcoin: A peer-to-peer electronic cash system.
\newblock 2008.

\bibitem{ozcelik2010business}
Yasin Ozcelik.
\newblock Do business process reengineering projects payoff? evidence from the
  united states.
\newblock {\em International Journal of Project Management}, 28(1):7--13, 2010.

\bibitem{peters2015trends}
Gareth Peters, Efstathios Panayi, and Ariane Chapelle.
\newblock Trends in cryptocurrencies and blockchain technologies: a monetary
  theory and regulation perspective.
\newblock 2015.

\bibitem{peters2016understanding}
Gareth~W Peters and Efstathios Panayi.
\newblock Understanding modern banking ledgers through blockchain technologies:
  Future of transaction processing and smart contracts on the internet of
  money.
\newblock In {\em Banking Beyond Banks and Money}, pages 239--278. Springer,
  2016.

\bibitem{spielman2016blockchain}
Avi Spielman.
\newblock {\em Blockchain: digitally rebuilding the real estate industry}.
\newblock PhD thesis, Massachusetts Institute of Technology, 2016.

\bibitem{szabo1997formalizing}
Nick Szabo.
\newblock Formalizing and securing relationships on public networks.
\newblock {\em First Monday}, 2(9), 1997.

\bibitem{underwood2016blockchain}
Sarah Underwood.
\newblock Blockchain beyond bitcoin.
\newblock {\em Communications of the ACM}, 59(11):15--17, 2016.

\bibitem{vanwersch2016critical}
Rob~JB Vanwersch, Khurram Shahzad, Irene Vanderfeesten, Kris Vanhaecht, Paul
  Grefen, Liliane Pintelon, Jan Mendling, Godefridus~G van Merode, and Hajo~A
  Reijers.
\newblock A critical evaluation and framework of business process improvement
  methods.
\newblock {\em Business \& Information Systems Engineering}, 58(1):43--53,
  2016.

\bibitem{weber2016untrusted}
Ingo Weber, Xiwei Xu, R{\'e}gis Riveret, Guido Governatori, Alexander
  Ponomarev, and Jan Mendling.
\newblock Untrusted business process monitoring and execution using blockchain.
\newblock In {\em International Conference on Business Process Management},
  pages 329--347. Springer, 2016.

\bibitem{yli2016current}
Jesse Yli-Huumo, Deokyoon Ko, Sujin Choi, Sooyong Park, and Kari Smolander.
\newblock Where is current research on blockchain technology?—a systematic
  review.
\newblock {\em PloS one}, 11(10), 2016.

\end{thebibliography}
